# Physics-informed Neural Network (PINN) to Predict Vibrational Stability of Inorganic Semiconductors

M. H. Zeb*, M. Z. Kabir

Department of Electrical and Computer Engineering

Concordia University

1455 Boul. de Maisonneuve Ouest, Montréal, Quebec H3G 1M8, Canada

*Email: mu_ze@live.concordia.ca, zahangir.kabir@concordia.ca

Keywords: Physics-informed neural network, Vibrational stability, Physics-based loss function, Born criteria

**Abstract**

We tackle the challenge of predicting vibrational stability in inorganic semiconductors for high-throughput screening, an essential attribute for evaluating synthesizability alongside thermodynamic stability, frequently missing in prominent materials databases. We create a physics-informed neural network (PINN) that incorporates the Born stability requirements directly into its loss function. This integration is a key learning constraint since it only allows the model to make predictions that do not violate fundamental physics. The model shows consistent and improved performance, having been trained on a dataset of 2112 inorganic materials with validated phonon spectra, and getting an F1-score of 0.83 for both stable and unstable classes. The model shows an AUC-ROC of 0.82 on a benchmark dataset of 1296 materials. Our PINN surpasses the best models in comparative tests, especially when it comes to accurately identifying unstable materials, which is crucial for a stability filter. This work offers a comprehensive screening tool for identifying materials and a methodology for incorporating domain knowledge to enhance predictive accuracy in materials informatics.

1. **Introduction**

There has been a remarkable increase in the use of machine learning (ML) across materials science, where ML has emerged as a key tool for rapid prediction of material properties and high-throughput screening.[1-5] A crucial initial filter in this screening process is material stability, which guarantees that computationally generated materials are synthetically feasible. This stability assessment primarily focuses on thermodynamic stability, as provided by major databases such as the Materials Project (MP), Open Quantum Materials Database (OQMD), and AFLOW.org, through metrics such as energy above the convex hull.[6-8] This method effectively eliminates materials susceptible to breakdown into competing phases; nevertheless, it neglects dynamical (vibrational) instability, a phenomenon wherein materials with energies even around the hull might exhibit instability due to soft phonon modes. [2,9-11]

The recognized method for investigating vibrational stability is the examination of phonon dispersion curves obtained from Density Functional Theory (DFT) calculations. DFT requires computations for large supercells to account for long-range interactions. So, these computations are computationally burdensome for high-throughput screening, a constraint that becomes even more pronounced for materials with large or complex unit cells.[12-15] Experimental approaches, such as inelastic neutron scattering and synchrotron X-ray spectroscopy, can yield phonon density of states; however, the expense of experimental equipment and the necessity of high-quality single crystals make it impractical, especially for high-throughput vibrational stability screening.[16-18]

As a result, a key ingredient, namely a vibrational stability filter, is missing from the materials discovery process. Promising candidates identified via machine learning for their target features may exhibit dynamic instability, resulting in an expensive and unproductive cycle of synthesis and characterization in the laboratory. This highlights the urgent need for the development of efficient and precise computational filters for vibrational stability, capable of operating at the high-throughput scale required by contemporary materials informatics.

The advancement of machine learning algorithms for predicting vibrational stability is still in its early stages. Several pioneering studies have sought to address this topic; nonetheless, their limitations underscore the complexity of the issue. A classifier was developed using data from around 3,000 materials, utilizing synthetic data augmentation to address class imbalance. [19] This classifier attained an Area Under the Curve (AUC) of 0.73 with the expanded data. This metric represents the area under the Receiver Operating Characteristic (ROC) curve, which plots the true positive rate against the false positive rate, thus quantifying overall classification performance. Despite good AUC value, its efficacy in accurately identifying unstable materials, the principal objective of a stability filter, was inadequate, a matter that will be statistically examined in a subsequent section.

Other works have been more narrowly focused in scope. For example, a model developed to accurately detect unstable two-dimensional (2D) materials identified unstable materials only 52% of the time.[20] Furthermore, its confinement to the realm of 2D materials constrains its applicability to the broader category of inorganic semiconductors. Concurrently, various methodologies have aimed to forecast the complete phonon dispersion curves utilizing advanced structures such as Euclidean neural networks (E3NN).[21] However, this was trained on, and hence effective for, only stable materials, without the necessary data on imaginary phonon modes crucial for identifying vibrational instability.

Acknowledging the constraints of exclusively data-driven methodologies, new approaches have tried to incorporate physical principles into model construction to enhance performance and generalizability.[22] One notable model is the Atomistic Line Graph Neural Network (ALIGNN), designed to forecast thermal and vibrational characteristics.[23] ALIGNN exhibited high precision in predicting parameters, including heat capacity and vibrational entropy. Nonetheless, its efficacy as a classifier for vibrational stability was inadequate, primarily due to the significant underrepresentation of unstable materials in the training data; approximately 15% of the phonon spectra exhibited imaginary modes, signifying instability. This continually underscores a fundamental obstacle: the inherent data imbalance between stable and unstable materials in current databases, which biases models toward overlooking the key indicators of instability. Overcoming this constraint requires an intelligent shift in approach. Initially, it necessitates a transition from passive dependence on existing databases to the proactive development of balanced datasets for model training. Secondly, and more importantly, we must utilize physics-informed modelling methodologies. Incorporating established physical constraints and relationships into the learning architecture enables the model to generalize more efficiently from limited and imbalanced data.[24] This strategy has proven highly effective in numerous scientific fields.[25-27]

By employing a physics-informed neural network architecture, this study effectively addresses the fundamental gaps previously identified in vibrational stability prediction, as evidenced by the superior performance of our model. Our key contribution is the direct integration of the Born stability criteria into the learning process via a customized supplementary loss function. The Born criteria constitute a required yet insufficient condition for complete dynamical stability, offering a robust physical precondition.[28,29] This integration enables the model to adhere to essential elastic constraints, significantly enhancing its predictive capabilities and generalizability. The details of the Born stability criteria and the loss function devised from this criterion will be discussed in the computational details section.

To facilitate a thorough assessment, we developed a benchmark dataset of 1296 inorganic semiconductors, carefully curated to guarantee a balanced representation of stable and unstable phases. Our model also demonstrated improved performance on this benchmark, as evidenced by a high Area Under the Receiver Operating Characteristic Curve (AUC-ROC) and additional classification metrics, which are further elaborated upon in the Results and Discussion section. This is, to our knowledge, the first benchmark dataset specifically designed for vibrational stability, establishing a crucial foundation for fair comparison and advancement of future models in this field. We also presented a critical evaluation of our model's shortcomings, examining instances of failure and suggesting possible remedies to inform future research in this emerging domain.

## 2. Computational details
### 2.1. Dataset creation

The efficacy of data-driven models, especially deep learning architectures, is fundamentally dependent on the quality and size of the given dataset. To address the established trade-off between data requirements and model complexity, physics-informed machine learning presents a robust framework by incorporating governing physical laws directly into the learning process.[27] This study operates within a hybrid framework, utilizing a curated materials dataset enhanced by the physics-based Born stability criterion. The preliminary dataset was derived from phonon dispersion curves acquired from the Materials Data Repository.[30] Dynamical stability was assessed from phonon dispersions, with negative frequencies (imaginary modes) denoting instability. Each material was assigned a binary State label, with 1 indicating a stable structure (absence of imaginary modes) and 0 indicating an unstable structure.[20] Additionally, a set of features was developed from the compositional and structural data of each material, chosen for their demonstrated effectiveness in predicting vibrational stability in inorganic semiconductors (see Supplementary Information for details).

To incorporate physics into our deep learning architecture, the "Born_Criteria" column was created. Details of the criteria are given in section 2.1.1. Materials that met the "Born stability criteria" were given a 1 (stable) label, while those that did not were given a 0 (unstable) label. Entries without proper State or "Born_Criteria" values were eliminated to maintain data integrity. The Born criteria were computed for 1545 materials due to the restricted availability of the elastic tensor. Of them, 1056 were stable materials, and 489 were unstable materials. To rectify the class imbalance, we employed a Random Forest Regressor (RFR) to forecast the born criteria of 567 materials, resulting in a final dataset of 2112 materials, evenly split between stable and unstable classifications.

The feature set was derived from all exclusively numeric columns, excluding compositions and the target variables (State and Born_Criteria). The dataset was cleaned by replacing infinite entries with NaN (not a number) and imputing missing feature values using column medians, ensuring robustness against outliers. All rows containing remaining NaN values after imputation were eliminated.

After balancing, all features were standardized using scikit-learn's preprocessing—StandardScaler module. The dataset was split into the training set (70%), the validation set (20%), and the test set (10%). The scaler was fit exclusively on the training data and subsequently applied to the validation and test sets, preventing data leakage and ensuring consistent processing. The fitted scaler was retained as a critical artifact for future predictions.

### 2.2. Model development

The core of the framework is a deep neural network defined in the model.py file, which can be accessed on GitHub via the link https://github.com/husnainzeb/pinn-vibrational-stability. We created a Multi-Layer Perceptron (MLP) particularly engineered for predicting material stability from tabular data. The network architecture was designed to represent intricate, non-linear interactions in the feature space while reducing overfitting. The architecture comprises an input layer, three hidden layers, and a single-node output layer for binary classification. A simplified representation of our architecture is shown in Figure 1.

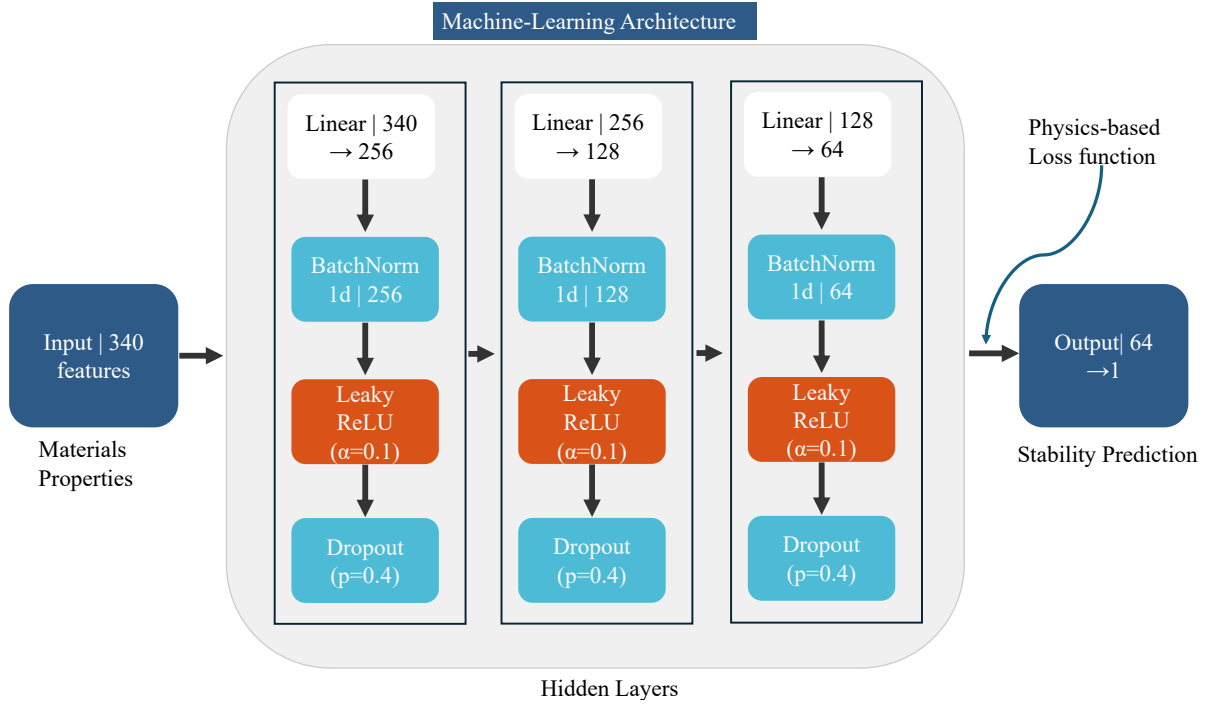

Figure 1: Simplified representation of the machine learning architecture developed in this work.

A key component of our methodology is a physics-informed training paradigm that incorporates domain-specific information directly into the learning process. We developed a hybrid loss function $\mathcal{L}_{physics}$ that integrates a conventional classification loss $\mathcal{L}_{prediction}$ with a customized physics-based penalty. This penalty term explicitly incorporates the Born stability requirements, imposing a penalty on the model when it incorrectly predicts a material as stable despite established elastic instability. This directs the model towards solutions that are both data-driven and physically consistent. The following equation gives the total loss function.[25]

$$\mathcal{L}_{total} = \mathcal{L}_{prediction} + \lambda \mathcal{L}_{physics} \qquad (1)$$

Where, $\mathcal{L}_{prediction}$ and $\mathcal{L}_{physics}$ are given by the following equations.

$$\mathcal{L}_{prediction} = \frac{1}{N}\sum_{i=1}^{N}[\alpha.(1-p_t)^\gamma . BCE(y_i, \hat{y}_i)] \quad (2)$$

$$\mathcal{L}_{physics} = \frac{1}{N}\sum_{i=1}^{N} I\{\hat{y}_i \geq 0.5 \land b_i = 0\} \quad (3)$$

N represents the batch size, $y_i$ is the true label for sample i, which can be 0 or 1, $\hat{y}_i$ is the predicted label and is equal to the sigmoid function $\sigma(.)$. More details about the sigmoid function are available in the supplementary file. In equation (3), $b_i$ represents the Born criteria output, and it is 0 if the material is unstable according to the Born criteria; it will be 1 if the Born criteria suggest that the material is stable. $\lambda$ is the penalty strength hyperparameter. BCE is the cross-entropy function and given by the following equation.

$$BCE(y_i, \hat{y}_i) = -[y_i \log(\hat{y}_i) + (1-y_i)\log(1-\hat{y}_i) \quad (4)$$

I{} is the indicator function and is defined as;

$$I\{A\} = \begin{cases} 1 & \text{if condtion A is true} \\ 0 & \text{if condition A is not true} \end{cases}$$

A baseline model with an identical architecture and preprocessing was trained using only the classification loss, excluding the physics-based constraint, to enable a rigorous comparison. This allows a direct assessment of the impact of our physics-informed methodology. See Supplementary Section S1 for full details of the network architecture, hyperparameters, and loss function.

*2.1.1. Born stability criteria*

We incorporate essential physics-based restrictions into our deep learning architecture by introducing a physics-informed loss function based on the Born elastic stability requirement. The criteria establish the essential conditions for a crystalline solid to remain stable under any homogenous, tiny strain, characterized by the positive definiteness of the elastic constant matrix.[28,31]

$$C_{ij} = \frac{1}{V_0}\left(\frac{\partial^2 E}{\partial \varepsilon_i \partial \varepsilon_j}\right) \qquad (5)$$

Here, $E$ is the energy of the crystal, $V_0$ is its equilibrium volume, $\varepsilon$ is the strain, and Cij represents the matrix of second-order elastic constants. Here, $i$ and $j$ run from 1 to 6, representing all the independent stress-strain relationships in the material system. This elastic matrix or stiffness matrix is a 6*6 symmetric matrix. There are 21 independent components, which get reduced even further based on symmetry. Complete details are given in the Supplementary file. The criteria state that a crystal is elastically stable only if its elastic stiffness matrix ($C$) is positive. This is equivalent to requiring all eigenvalues or all leading principal minors of $C$ to be positive. While these are the necessary and sufficient conditions, several weaker, necessary-but-insufficient conditions exist, such as $C_{ii} > 0$ and $(C_{ij})^2 < C_{ii}C_{jj}$ for all $I\ j$. The complete set of conditions used to construct the loss function for each of the crystal systems is given in the supplementary file. Our loss function penalizes model predictions that designate a crystal as stable if its elastic constants violate the Born criteria for its specific crystal system.

It is critical to recognize the scope and limitations of this approach. The Born criteria are a subset of the broader requirement for full dynamical (vibrational) stability. A crystal that is dynamically stable across the entire Brillouin zone, which means all phonon modes have positive frequencies, will necessarily satisfy the Born criteria, as these govern the long-wavelength acoustic phonon branches. However, the converse is not true: a crystal fulfilling the Born criteria can still be dynamically unstable due to soft optical phonon modes at finite wavevectors, which the elastic constants do not capture. Thus, the Born criteria represent a necessary but not sufficient condition for overall vibrational stability. The schematic given in Figure 2 explains this relationship.

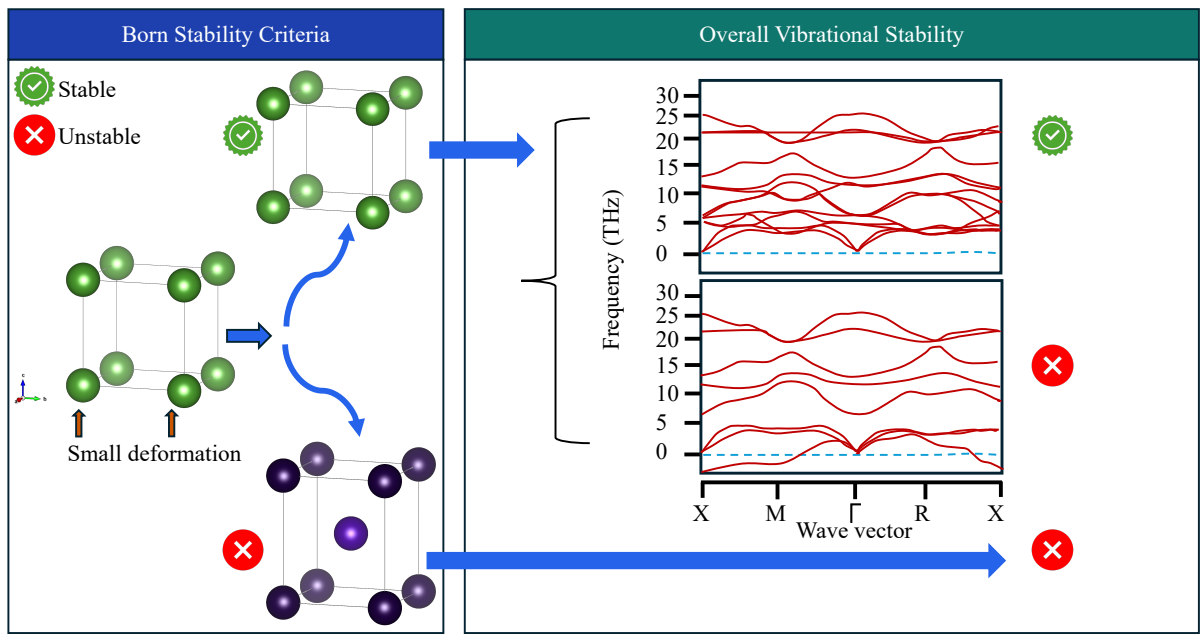

Figure 2: Relationship between Born stability criteria and overall vibrational stability.

This limitation restricts the Born criteria to the stability assessment of long-wavelength acoustic branches; yet their incorporation as a soft physical constraint offers a significant learning bias for our machine learning model. When combined with a data-driven loss, this hybrid approach guides the learning process towards candidate materials that are, at a minimum, satisfying a necessary but not sufficient condition for vibrational stability. The performance improvement is demonstrated in Table 1, comparing models with and without the physics loss, which validates this strategy. This success aligns with the established paradigm that couples data-driven models with incomplete yet relevant physical theories, which can significantly enhance predictive performance and generalization. We will further explore the limitations and potential extensions of this physics-informed approach in the Results and Discussion section.

## 3. Results and discussion

The developed Physics-Informed Neural Network (PINN) exhibits strong and exceptional capabilities in categorizing the vibrational stability of inorganic semiconductors, utilizing physical principles to improve predictive performance. The results presented in Table 1 indicate that our model demonstrates strong and equitable performance across all stability classes. It achieves F1 score of 0.83 across both classes. The incorporation of the physics-based loss function significantly influences this performance, as demonstrated by the steady enhancement in metrics relative to the model that was trained without this physical constraint.

A key advantage of our study lies in the thorough validation of our model's generalizability using an independent benchmark dataset, which is an essential measure for demonstrating real-world applicability that was significantly lacking in all similar previous research. The findings, illustrated in Table 2, validate the model's strong predictive capabilities, showcasing F1-scores of 0.82 for the stable class and 0.81 for the unstable class. The overall discriminative capability is highlighted by a high Area Under the Curve (AUC) of 0.82 for the Receiver Operating Characteristic (ROC) for the benchmark dataset (Figure 3, green line) and an AUC of 0.83 for the test set (Figure 3, red line). A comparative analysis with recent advanced models, as outlined in Table 3, underscores the unique benefits and wider applicability of our approach. Models such as ALIGNN demonstrate a commendable recall for stable compounds at 0.95; however, they show a significant limitation in detecting unstable materials, with a U-F1 score of 0.66. This reflects a direct consequence of their significantly imbalanced dataset (stable: unstable = 4.9:1) and the absence of independent benchmark testing, raising concerns about their generalizability for a balanced classification task. In the same way, the RFR (Random Forest Regressor) developed encounters with challenges with the unstable class (U-F1: 0.63).

Table 1: Classification metrics for the Physics-informed neural network developed in this work.

| Class | Precision | Recall | F1 |
| --- | --- | --- | --- |
| Unstable | 0.82 | 0.84 | 0.83 |
| Unstable (without physics-based loss function) | 0.81 | 0.74 | 0.77 |
| Stable | 0.84 | 0.82 | 0.83 |
| Stable (without physics-based loss function) | 0.76 | 0.82 | 0.79 |

Contextualizing the scope and input data of these models is crucial. The XGBoost (Extreme Gradient Boosting) developed in another study presents a high AUC of 0.90; however, it is only for 2D materials and depends solely on phonon calculations derived from three high-symmetry points.[20] This simplification could undermine the model's accuracy and undoubtedly limits its applicability to 3D bulk systems. In contrast, our dataset is constructed from complete phonon dispersion curves, offering a more rigorous and physically comprehensive foundation for stability assessment. Moreover, various studies have integrated physics via architectural inductive biases. For example, Chen et al. (E3NN, 2021) encoding translation, rotation and inversion symmetry or ALIGNN representing local bond angles, their main emphasis was not on the explicit binary classification of vibrational stability. The E3NN work focused mainly on different thermal properties, and neither model underwent the same level of rigorous validation for this critical task as our PINN does. The model integrates physics directly into the learning objective through the loss function and clearly outperforms these architectures for the specific task of stability classification.

Figure 4 illustrates the Frequency vs wave vector plots of three distinct unstable materials together with their respective crystal systems. The predicted and actual stability statuses are also displayed. Our model effectively identified the imaginary frequencies linked to both long-range and short-range wavelengths in the cases of ZnS and KICl$_2$. Table 4 presents ten entries for unstable materials from our test dataset. The entire collection is accessible on our GitHub repository. (https://github.com/husnainzeb/pinn-vibrational-stability)

Table 2: Classification metrics for the benchmark dataset.

| Class | Precision | Recall | F1 |
| --- | --- | --- | --- |
| Unstable | 0.80 | 0.82 | 0.81 |
| Stable | 0.83 | 0.81 | 0.82 |

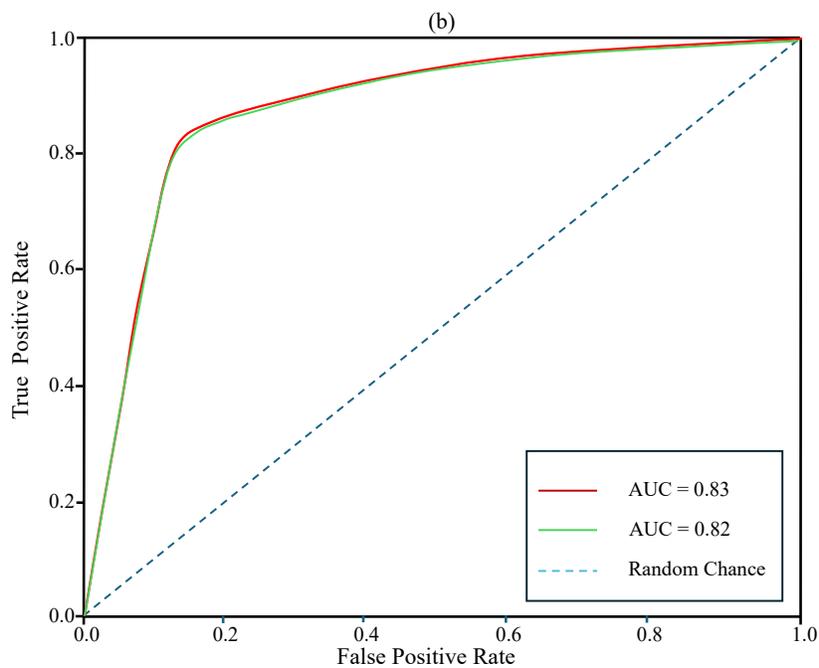

Figure 3: The receiver operating curve for the test set (red) and the benchmark dataset (Green).

Our framework ultimately delivers exceptional and balanced performance by employing a more streamlined and easily computable set of features that are directly derived from material composition and structure. This approach circumvents the challenges associated with creating extremely high-dimensional feature sets or training intricate graph neural networks on extensive datasets. Moreover, we used a dropout of 0.4 in our neural network architecture, which randomly switches off 40% of the neurons every time and helps the model to overcome the overfitting problem, which makes it less sensitive to noise and more likely to capture the underlying physical principles of stability rather than spurious correlations in the dataset. This design, combined with the physics-based loss function, directly contributes to our model's superior performance on the independent benchmark test, demonstrating its ability to generalize beyond its initial training distribution.

Table 3: Comparison of our work with existing literature.

|  | RFR (2023) | E3NN (2021) | ALIGNN (2023) | XGBoost (2023) | This work (2025) |
|---|---|---|---|---|---|
| Dataset Size | 3100 | 1500 | 14000 | 3212 | 2112 |
| Recall | S:0.79 U:0.68 | - | S: 0.95 U: 0.61 | - | S:0.82 U:0.84 |
| F1 Score | S:0.81 U:0.63 | - | S: 0.93 U: 0.66 | - | S:0.83 U:0.83 |
| Stable: Unstable | 2.2: 1 | - | 4.9: 1 | 1: 1.1 | 1: 1 |
| AUC | 0.73 | - | - | 0.90 | 0.83 |
| Benchmark Dataset | ❌ | ❌ | ❌ | ✅ | ✅ |
| Physics-based | ❌ | ~ | ~ | ❌ | ✅ |

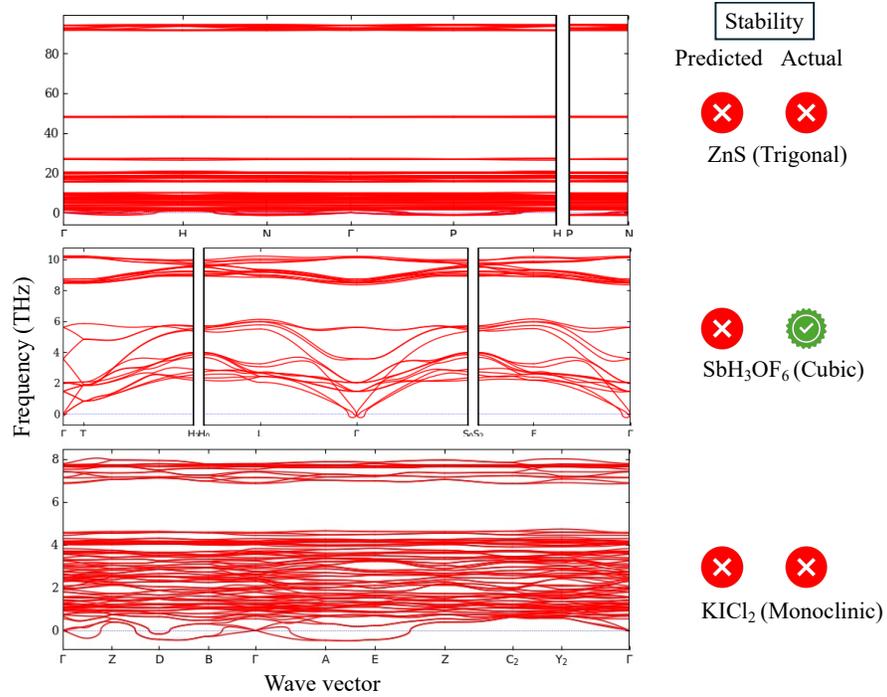

Figure 4: Frequency vs wave vector for three materials from our test set. The crystal system, actual and predicted stability statuses, is also shown. The presence of frequencies below zero shows vibrational instability.[32]

Table 4: Results for some of the unstable materials from our test set. The Complete set is available in our GitHub repository.

| Materials-ID | Composition | Crystal system | True state | Predicted state |
|---|---|---|---|---|
| mp-22023 | $Pb_3SO_6$ | Monoclinic | 0 | 0 |
| mp-644419 | LiHS | Orthorhombic | 0 | 0 |
| mp-643579 | $SbH_3OF_6$ | Cubic | 0 | 0 |
| mp-554405 | ZnS | Trigonal | 0 | 1 |
| mp-628908 | $KICl_2$ | Monoclinic | 0 | 0 |
| mp-16993 | $Er_2SiO_5$ | Monoclinic | 0 | 0 |
| mp-20025 | $Mn(GaS_2)_2$ | Tetragonal | 0 | 0 |
| mp-757777 | $LiSiBi_3O_7$ | Hexagonal | 0 | 0 |
| mp-554463 | $Na_3LuSi_2O_7$ | Hexagonal | 0 | 0 |
| mp-4867 | $CaAl_4O_7$ | Monoclinic | 0 | 0 |

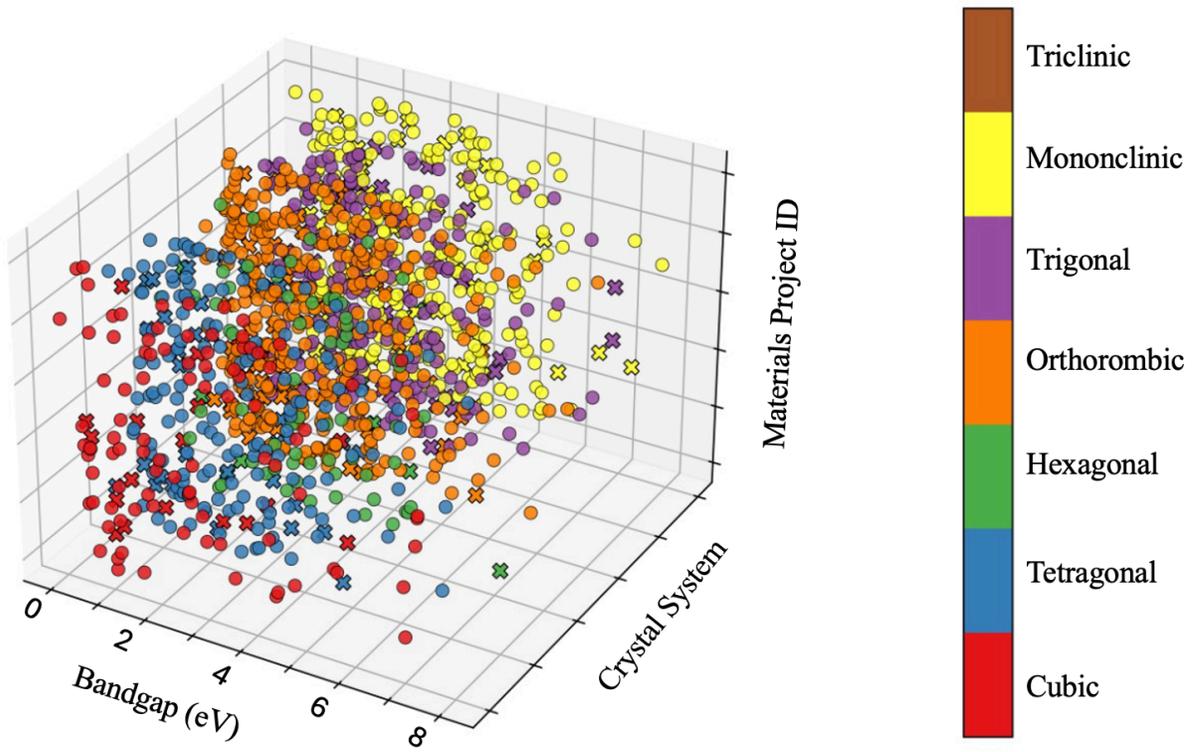

Figure 5: Visual representation of materials from our benchmark dataset showing spread across different crystal systems and bandgaps.

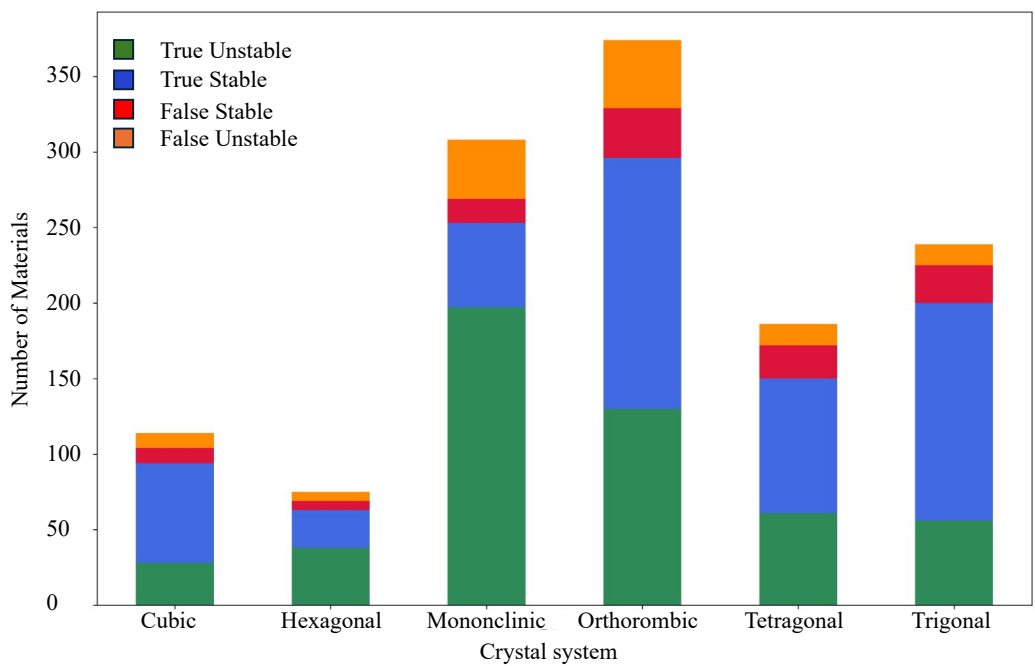

Figure 6: Visual representation of the performance of our model across different crystal systems in our benchmark dataset.

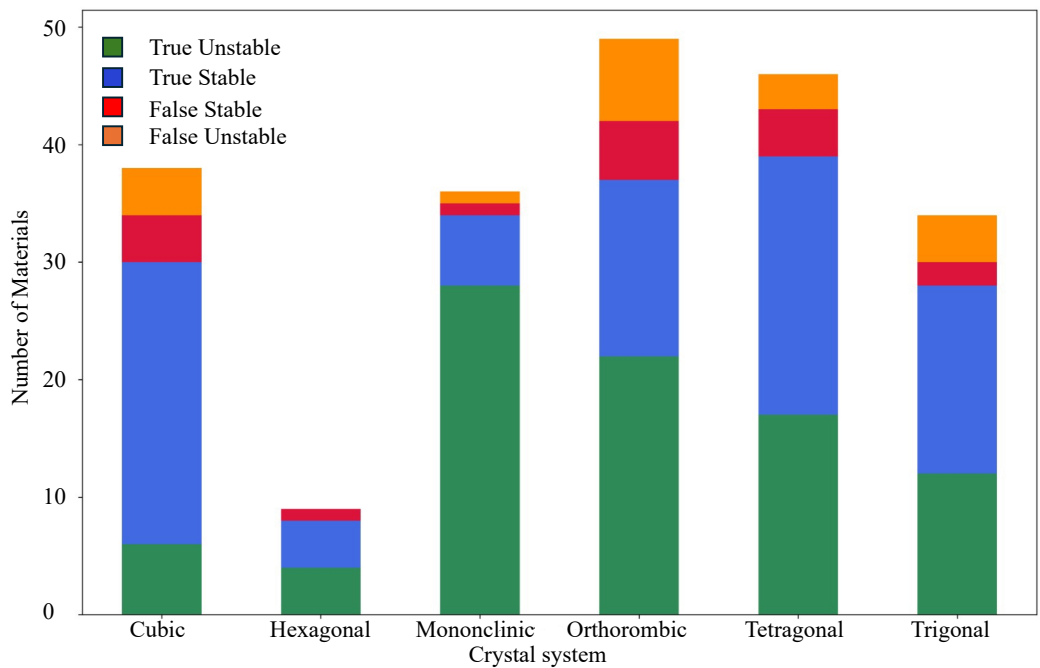

Figure 7: Visual representation of the performance of our model across different crystal systems in our test dataset.

In short, through the direct integration of physical domain knowledge into the learning objective, the utilization of a comprehensive physical dataset, and the crucial step of validating performance on independent benchmark data, we have created a model that surpasses existing methods in accuracy, robustness, generalizability, and practical reliability for real-world materials screening. Figure 5 provides a visual representation illustrating the performance of our model across various crystal systems and a broad spectrum of band gaps.

Despite its impressive performance, our model has limitations that point to future research directions. Figures 6 and 7 show the number of materials for different crystal systems in our benchmark and test datasets, respectively. We were unable to find materials from the triclinic crystal system in the MDR Phonon database. Inclusion of those materials can improve the generalizability of the machine learning model. Moreover, the current physics-based loss, rooted in the Born criteria, is most effective for high-symmetry crystals; its utility diminishes for low-symmetry structures where the criteria are less restrictive. This suggests a compelling opportunity to explore alternate or supplementary physical laws, such as those governing lattice anharmonicity or specific soft-mode instabilities, which can help to create a more universally robust classifier. Addressing these points is beyond the scope of the present study and will be the focus of our subsequent efforts.

## 4. Conclusion

In this work, we have presented a physics-informed machine learning framework that sets a new benchmark for the high-throughput prediction of vibrational stability in 3D inorganic semiconductors. Integrating the Born stability criteria into the neural network's loss function enables the model to learn from both data and fundamental physical laws. This strategy is effective, resulting in a classifier that demonstrates outstanding and equitable performance in both stable and unstable classes, which is essential for a functional screening tool.

The superiority of our model is clearly demonstrated through rigorous benchmarking against existing methods. This model attains the highest documented F1-score for the difficult unstable class, while also demonstrating superior performance for stable materials, surpassing models that depend on considerably larger feature sets or more intricate architectures. By employing a balanced, independent benchmark for validation, this work addresses a key methodological gap in the field, providing strong evidence for the model's robustness and generalizability.

This study highlights a crucial principle: the incorporation of incomplete physical knowledge can significantly improve machine learning models, guiding them away from misleading correlations and towards meaningful physical generalizations. The availability of a robust, practical, and rigorously validated tool, along with a foundational benchmark for future research, addresses a significant barrier in the computational materials discovery pipeline, facilitating the more reliable identification of synthesizable, next-generation materials.